\documentclass{aastex631}
\usepackage[utf8]{inputenc}
\usepackage{CJKutf8}
\usepackage{graphicx}
\usepackage{latexsym}
\usepackage{amssymb}
\usepackage{epsf}
\usepackage{amsmath}
\usepackage{booktabs}
\usepackage{gensymb}
\usepackage{threeparttable} 
\usepackage{tabularx}

\begin{document}
	\begin{CJK}{UTF8}{gbsn}
		
		\title{Exploring the Wolf Approach to Constraining NIR Extinction Laws in the Corona Australis Molecular Cloud}
		
		\author[0009-0007-2880-849X]{Botao Jiang (江博韬)}
		\affiliation{Center for Astrophysics, Guangzhou University, Guangzhou 510006, People's Republic of China}
		
		\author[0000-0001-9328-4302]{Jun Li (李军)}
		\affiliation{Center for Astrophysics, Guangzhou University, Guangzhou 510006, People's Republic of China}
		
		\author[0000-0002-5435-925X]{Xi Chen (陈曦)}
		\affiliation{Center for Astrophysics, Guangzhou University, Guangzhou 510006, People's Republic of China}
		
		\correspondingauthor{Jun Li, Xi Chen}
		\email{lijun@gzhu.edu.cn, chenxi@gzhu.edu.cn}
		

	\begin{abstract}
		The viability of the star count (Wolf) method is assessed as a means of constraining the near-infrared (NIR) extinction law toward the Corona Australis molecular cloud. Using deep $JHK_S$ photometry from the VISIONS survey, extinction maps with 1 arcmin spatial resolution are constructed. The derived extinction ratios are $A_J/A_H=1.73\pm0.07$, $A_H/A_{K_S}=1.70\pm0.11$, and $A_J/A_{K_S}=3.02\pm0.22$, which are consistent with Galactic literature means. Assuming a power-law form ($A_\lambda \propto \lambda^{-\alpha}$) for the NIR extinction law, we derive indices of $\alpha\approx 2.0$ across all wavelength combinations, with no statistically significant wavelength dependence throughout the NIR wavelength range. While spatial variations in extinction properties are tentatively observed across the cloud, concerns persist regarding the impact of photometric completeness, and the role of reference field selection. Continued research is required to refine the approach, and scrutinize the veracity of potential extinction law variations over a more expansive region of sky.
	\end{abstract}
	
	
	\section{Introduction} \label{sec:intro}
    
	The interstellar extinction law characterizes how dust grains in the interstellar medium (ISM) scatter and absorb light as a function of wavelength \citep[e.g.,][]{Cardelli1989,Draine2003}. In molecular clouds, where optical radiation is heavily attenuated by dust, near-infrared (NIR) observations become particularly valuable due to their ability to probe deeper into molecular clouds. Accurate quantification of the NIR extinction law facilitates the construction of high-precision extinction maps, which are fundamental for examining the structure, density, and mass distribution of molecular clouds \citep[e.g.,][]{Lombardi2001,Lombardi2011}. These maps reveal the detailed morphology of star-forming regions, enabling the identification of embedded stellar populations and the determination of gas and dust column densities that provide critical constraints on the initial conditions of star formation. Moreover, precise characterization of the NIR extinction law offers insights into the physical properties of dust grains, including their composition, size distribution, and alignment in the ISM \citep[e.g.,][]{Weingartner2001,Indebetouw2005}.
	
	Extinction law has been extensively studied in the ultraviolet and optical regimes, where it can be characterized by the total-to-selective extinction ratio $R_V$, defined as the ratio of optical extinction $A_V$ to color excess $E(B-V)$. The average $R_V$ value for the diffuse ISM in the Milky Way is 3.1, with values ranging from 2.0 to 6.0 depending on the line of sight \citep{Cardelli1989,Fitzpatrick1999}. NIR extinction is notably less sensitive to variations in $R_V$, making it a more reliable tracer of total column density across diverse environments \citep{Cardelli1989}. The NIR extinction law is commonly parameterized using a power-law form $A_\lambda \propto \lambda^{-\alpha}$ \citep{Martin1990}. However, the reported values of the extinction index $\alpha$ have changed significantly over recent decades, from approximately 1.5 \citep[e.g.,][]{Rieke1985,Draine1989} to values exceeding 2.0 or even reaching 2.5 \citep[e.g.,][]{Nishiyama2006,Stead2009,Gosling2009,schodel2010,Fritz2011,Alonso-Garcia2017,Nogueras-Lara2018}. Beyond this discrepancy, recent studies have reported evidence suggesting possible variations in the extinction index between different NIR bands, specifically between $JH$ and $HK_S$ \citep{Nogueras-Lara2018,Nogueras-Lara2019}. These uncertainties in the NIR extinction law may introduce significant challenges in the construction of reliable NIR extinction maps.
	
	Previous measurements of NIR extinction law have predominantly focused on the Galactic center or Galactic plane \citep[e.g.,][]{Nishiyama2009,Nogueras-Lara2019,Sanders2022}, or have derived Galactic average values \citep[e.g.,][]{Wang2014,Stead2009}. Systematic discrepancies emerge from different observational techniques and stellar population tracers employed across these studies, potentially yielding results that represent averaged properties across diverse interstellar environments. Within dense molecular clouds, dust grains undergo significant evolution through coagulation and ice mantle formation, altering their size distribution and structure, which in turn influences the observed NIR extinction properties \citep{Jones2013,Ormel2011,Cao2024,Li2024a}. Furthermore, observational depth limitations significantly constrain our ability to probe dust properties within the densest molecular cloud regions, where extinction gradients are steepest \citep{Olofsson2010,Foster2013}. Particularly challenging in high-extinction environments is the degeneracy between extinction effects and filter transmission profiles, which introduces wavelength-dependent biases that can substantially impact extinction law determinations when applying standard photometric techniques \citep{MaizApellaniz2020,Wang2019}. This complex interplay of environmental variations and methodological limitations necessitates carefully controlled comparative studies to establish a comprehensive understanding of NIR extinction law.
    
	In this work, we assess the viability of the star count (Wolf) method as a means of constraining the NIR extinction law toward the Corona Australis (hereafter CrA) molecular cloud using deep photometric data from the VISIONS survey. Section \ref{sec:da} describes the data and quality control. Section \ref{sec:Met} presents the method. Section \ref{sec:ReDis} describes the results and discussion. Finally, we summarize the main conclusions in Section \ref{sec:Sum}. 
	
	\section{data}\label{sec:da}
	
	\subsection{The Corona Australis Molecular Cloud}
	CrA is a relatively isolated molecular cloud complex located in the southern sky at high Galactic latitude ($l\approx$ $359\degree-360\degree$, $b\approx$ $-17\degree$ to -20$\degree$) at a distance of approximately 150\,pc \citep{Zucker2020}. This region represents an active star-forming region characterized by complex filamentary structures and several dense cores, with a total estimated mass of $\sim 7000\,M_\odot$ \citep{Alves2014,Bresnahan2018}. The central region hosts a rich population of young stellar objects, Herbig-Haro objects, molecular outflows, and prestellar cores \citep{Peterson2011}. Observations based on 2MASS data reveal that CrA exhibits significant internal extinction variations, with visual extinction ($A_V$) reaching 25\text{--}30\,mag in its densest core regions while maintaining moderate extinctions of  $A_V\approx$ 1\text{--}3\,mag in peripheral areas \citep{Alves2014}. The relatively isolated structure of CrA minimizes contamination from background or foreground clouds, making it an ideal laboratory for studying extinction properties within a well-defined molecular environment.
	
	\subsection{Data and Quality Control}
	
	The NIR photometric data analyzed in this study are from the Visible and Infrared Survey Telescope for Astronomy (VISTA) Star Formation Atlas (VISIONS) \citep{Meingast2023b,Meingast2023a}, specifically from observations of the CrA molecular cloud complex, accessible through the ESO Archive Science Portal \footnote{https://archive.eso.org/scienceportal/home?data\_collection=VISIONS\&publ\_date=2022-02-28}. VISIONS is a public survey conducted with the VISTA Infrared Camera (VIRCAM), provides imaging in the $J$ (1.25\,$\mu$m), $H$ (1.65\,$\mu$m), and $K_S$ (2.15\,$\mu$m) bands. We utilize the deep observations from Data Release 2, which cover approximately 1.8\,square degrees targeting the regions of highest column density. The photometric calibration, performed against 2MASS, yields negligible residual color terms, while the astrometric solution, calibrated against Gaia DR3, achieves precisions of $\sim$10\,milliarcseconds for all sources and $\sim$3\,milliarcseconds for bright sources.
	
\begin{figure}[ht!]
		\includegraphics[width=\textwidth]{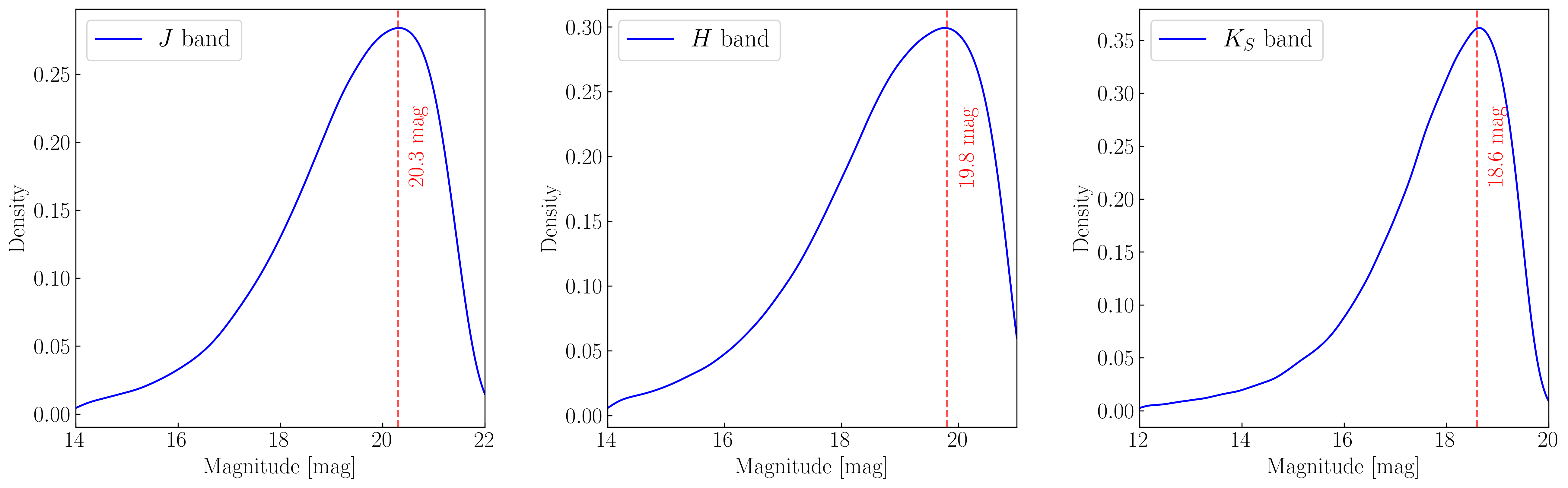}
		\caption{Density plots of photometric data of the CrA molecular cloud in the $J$, $H$, and $K_S$ bands from the VISIONS.The dashed lines in all the panels indicate the completeness limiting magnitudes of 20.3\,mag, 19.8\,mag, and 18.6\,mag in the $J$, $H$, and $K_S$ bands, respectively.
			\label{fig:limit_mag}}
	\end{figure}

	To ensure robust source selection and high photometric quality, we adopt strict criteria: photometric uncertainties in all bands ($J$, $H$, and $K_S$) are smaller than 0.4\,mag, and ELLIPTICITY values are below 0.3 to minimize contamination from background galaxies and image artifacts. To assess the completeness limits of our photometric catalogs, we employ the histogram turnover method \citep{Damian2021}, which generally defines the 90\% completeness level. From the $JHK_S$ magnitude distributions shown in Figure \ref{fig:limit_mag}, the completeness limits are determined to be $J \sim 20.3$, $H \sim 19.8$, and $K_S \sim 18.6$ mag. These completeness thresholds are subsequently applied in the construction of the extinction maps. 
	
	\section{Method}\label{sec:Met}
	\subsection{Wolf Method}\label{sec:ST}
	We employ the Wolf method (also called star count method) to construct extinction maps in the $J$, $H$, and $K_S$ bands for the CrA region. This technique, first introduced by \cite{Wolf1923} and subsequently widely applied in extinction mapping \citep{Cambresy1999,Dobashi2005}, is based on the principle that observed stellar density decreases with increasing extinction due to reduced detection of fainter stars. The method involves partitioning the observed region into a defined grid and comparing the stellar counts in each cell with those from a reference field. This approach assumes that all stars lie behind the cloud and that the stellar population remains uniform across the field.
	
\begin{figure}[ht!]
		\includegraphics[width=\textwidth]{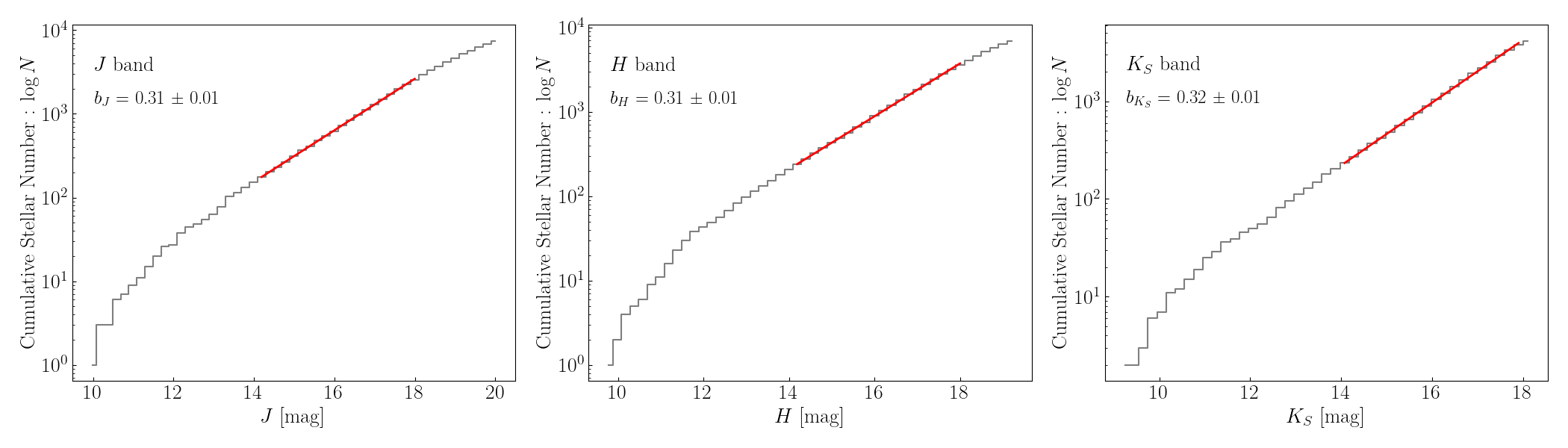}
		\caption{Wolf diagram showing the logarithmic cumulative frequency measured in the low-extinction region (284.84$\degree$ $\leq$ \textbf{RA} $\leq$ 285.1$\degree$, $-$36.5$\degree$ $\leq$ Dec $\leq$ $-$36.25$\degree$). Bin size is 0.2\,mag. The red lines indicate the best-fit lines using Equation \ref{cumulative_fd}. Slopes of $b_J = b_H = 0.31 \pm 0.01$\,mag$^{-1}$ for the $J$ and $H$ bands, and $b_{K_S} = 0.32 \pm 0.01$ \,mag$^{-1}$ for the $K_S$ band are derived, all measured within the magnitude range of 14.0--18.0\,mag.}
		\label{cumulative}
	\end{figure}

	For each pixel in our extinction map, we calculate the extinction using:
	\begin{equation}
		A_\lambda = \frac{1}{b_\lambda} \log\left(\frac{D_{\text{ref}}}{D_{\text{cloud}}}\right),
	\end{equation}
	where $A_\lambda$ represents the extinction in band $\lambda$ ($J$, $H$, or $K_S$), $D_{\text{ref}}$ and $D_{\text{cloud}}$ denote the stellar densities in the reference field and cloud region, respectively. Typically, $D_{\text{ref}}$ varies with Galactic coordinates, often described as an exponential function of the Galactic latitude \citep{Dobashi2005}. However, given the relatively small area of the coverage in this study, we assume that the background stellar density remains constant. $b_\lambda$ is the slope of the stellar luminosity function in the considered band. We determine $b_\lambda$ empirically from the reference field data using a cumulative frequency diagram of star number \citep[Wolf diagram;][]{Wolf1923}, described by:
	
	\begin{equation}\label{cumulative_fd}
		\log N = b_{\lambda}m_{\lambda} + C_1,
	\end{equation}
	where $m_{\lambda}$ represents the magnitude in band $\lambda$ and $C_1$ is a constant. To quantify systematic uncertainties arising from reference field selection, we analyzed 10 randomly selected reference fields across the survey area. Figure \ref{cumulative} shows the Wolf diagram for a representative reference region ($284.84\degree \leq RA \leq 285.1\degree$, $-36.5\degree\leq$ Dec $\leq-36.25\degree$), yielding best-fit values of $b_{\lambda}$ = $0.31 \pm 0.01$ mag$^{-1}$, $0.31 \pm 0.01$ mag$^{-1}$, and $0.32 \pm 0.01$ mag$^{-1}$ for the $J$, $H$, and $K_S$ bands, respectively, determined over the magnitude range $14.0 \leq m_{\lambda} \leq 18.0$ mag. Analysis of all ten reference fields reveals remarkably consistent behavior, with mean $b_{\lambda}$ values of 0.31, 0.31, and 0.32 for the three bands. The field-to-field standard deviation of 0.01 is comparable to the individual fitting uncertainty of 0.01, indicating that both statistical and systematic uncertainties contribute similarly. We therefore propagate these uncertainties in quadrature to obtain the total uncertainty of $\sigma_{b_\lambda}\sim 0.014$ for all three bands.
	
	We construct our extinction maps using a regular grid with 1\,arcmin$^2$ pixels. To optimize between spatial resolution and noise reduction, we calculate stellar densities using a Gaussian smoothing kernel of 2\,arcmin.
	
	The statistical uncertainty in the extinction measurements follows:
	\begin{equation}
		\sigma_{A_\lambda} = \frac{\log_{10}{e}}{b_{\lambda}}\sqrt{\frac{1}{N}+\frac{1}{N_{\text{refer}}}+\left(\frac{A_\lambda}{\log_{10}{e}}\sigma_{b_\lambda}\right)^2},
	\end{equation}
	where $N$ and $N_{\text{refer}}$ represent the number of stars within a pixel and the average number of stars within a reference field of view of a pixel, respectively, $\sigma_{b_\lambda}$ is the uncertainty of $b_\lambda$ as described above.

	\subsection{Extinction Maps of $A_J$, $A_H$ and $A_{K_S}$}
	
\begin{figure}[ht!]
		\includegraphics[width=0.5\textwidth]{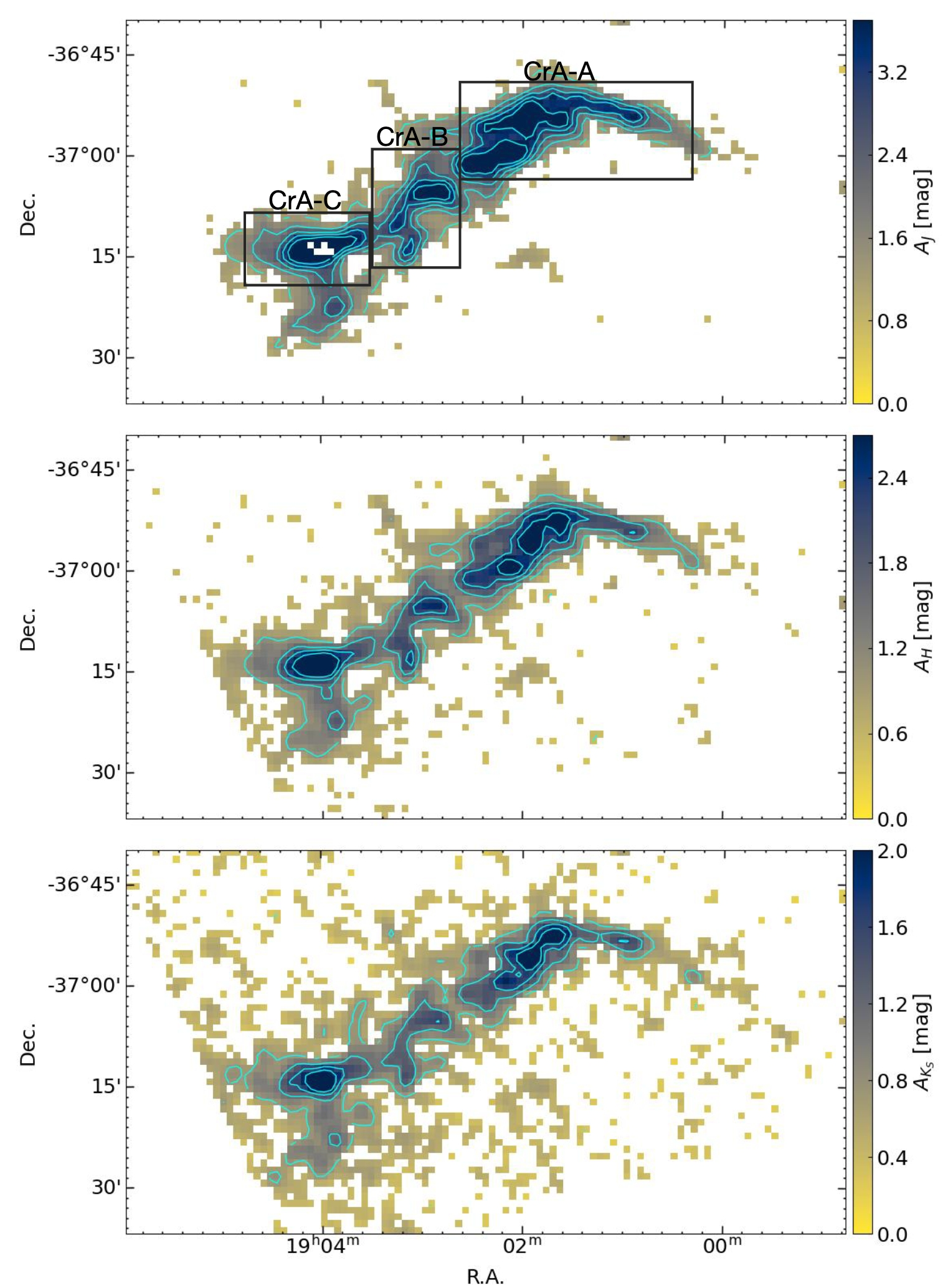}
		\centering
		\caption{
			Extinction maps of the CrA molecular cloud derived from VISIONS survey data, shown in $J$ (top), $H$ (middle), and $K_S$ (bottom) bands. Contour levels in the top panel begin at $A_J = 1.2$\,mag with increments of 0.5\,mag. The middle panel displays $A_H$ contours starting at 0.5\,mag with 0.4\,mag intervals, while the bottom panel shows $A_{K_S}$ contours beginning at 0.3\,mag with 0.4\,mag step sizes.
			\label{fig:extinction_map}}
	\end{figure}

	Figure \ref{fig:extinction_map} presents the derived extinction maps in the $J$, $H$, and $K_S$ bands constructed using the star count method detailed in Section \ref{sec:ST}. White regions indicate areas where reliable measurements are unattainable due to insufficient data or sub-threshold extinction values. The maps reveal maximum values of 6.8, 4.1, and 2.7\,mag in $A_J$, $A_H$, and $A_{K_S}$, respectively, with characteristic wavelength-dependent extinction patterns. In the densest central regions, the $A_J$ map exhibits greater saturation than the $A_{K_S}$ map, suggesting potential underestimation of extinction in heavily obscured areas due to the limited detectability of background stars at shorter wavelengths. While the $A_J$ map effectively traces moderate extinction regions ($A_J\sim 1.2$\,mag), the $A_{K_S}$ map predominantly highlights the densest cloud structures.
	
	The extinction maps demonstrate typical noise levels of $\sigma_{A_{J}}$ = 0.44\,mag, $\sigma_{A_{H}}$ = 0.41\,mag, and $\sigma_{A_{K_S}}$ = 0.49\,mag. With reference stellar densities of approximately 30 stars per pixel in the $J$ and $H$ bands and 17 stars per pixel in the $K_S$ band, extinction uncertainties are primarily governed by Poisson statistics, proportional to $\sqrt{N}$ of the observed stellar density. Consequently, measurement reliability decreases in high-extinction regions where background stars are severely attenuated. At our chosen resolution (1\,arcmin pixel scale, 2\,arcmin smoothing kernel), the maps resolve structures at approximately 0.05\text{--}0.1\,pc scale, sufficient for identifying individual dense cores and filamentary structures potentially associated with star formation processes.

	\section{Results and Discussion}\label{sec:ReDis}
	
	\subsection{Determination of reddening slopes in the NIR}\label{sec:reddening_slope}

	The NIR extinction maps derived from Section \ref{sec:Met} provide a valuable foundation for quantifying the NIR dust properties in the CrA cloud. These complementary datasets enable characterization of the wavelength-dependent extinction via direct measurements of relative extinction between bands (i.e., $A_J/A_H$, $A_H/A_{K_S}$, and $A_J/A_{K_S}$). By comparing these independent measurements across our mapped region, we can establish robust constraints on the NIR extinction law and its potential spatial variations.
	
\begin{figure}[ht!]
		\includegraphics[width=\textwidth]{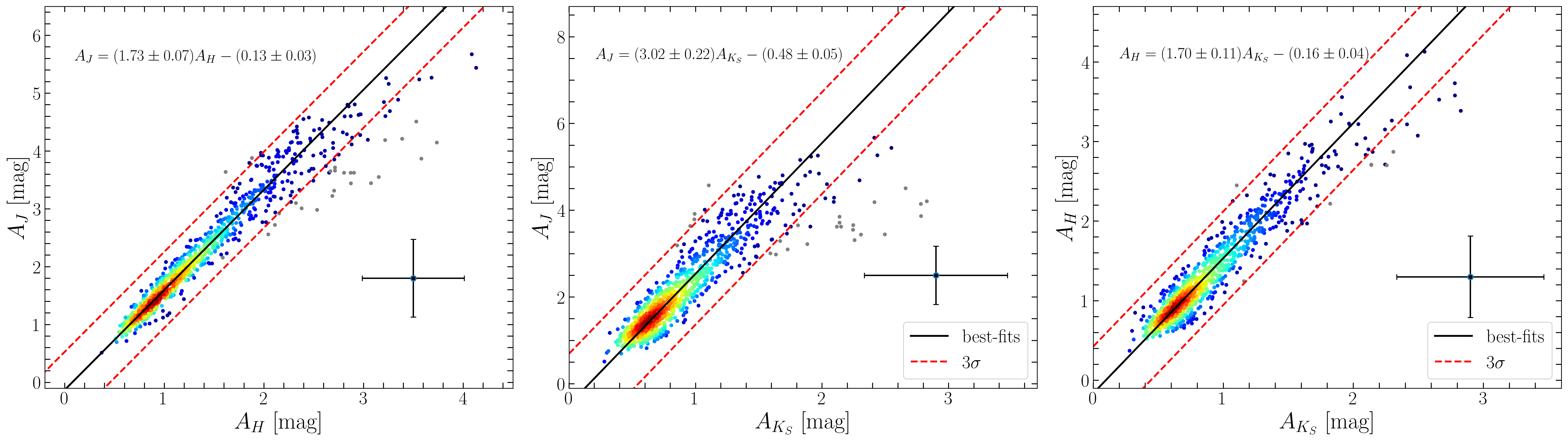}
		\caption{Correlations between extinction values in different NIR bands ($A_J$ vs. $A_H$, $A_J$ vs. $A_{K_S}$, and $A_H$ vs. $A_{K_S}$) for the CrA molecular cloud observed by the VISIONS survey. Each panel explicitly reports the fitted slope coefficient and intercept (zeropoint) with associated uncertainties.} The color scale indicates the density of data points. Black solid lines represent the best linear fits after removing $3\sigma$ outliers, with fit parameters annotated in each panel. Red dashed lines denote the $3\sigma$ boundaries, and points identified as outliers are shown in grey. Error bars in each panel represent the characteristic noise levels in the extinction measurements.
			\label{fig:Ax_Ay}
	\end{figure}

	Figure \ref{fig:Ax_Ay} presents the correlations between extinction values in the $J$, $H$, and $K_S$ bands across the CrA molecular cloud region. After excluding data points with negative extinction values, we observe robust linear relationships between $A_J$ and $A_H$, $A_H$ and $A_{K_S}$, and $A_J$ and $A_{K_S}$. The scatter in these relationships remains relatively small for moderate extinction values but increases noticeably for $A_H > 2.5$\,mag or $A_{K_S} > 1.5$\,mag. This increased dispersion likely results from the lower stellar density in high-extinction core regions, which may introduce larger measurement errors or lead to underestimation of extinction values.

    Through linear fitting of extinction between wavebands, we directly determined the extinction ratios. To quantify these relationships, we employed the \texttt{LTS\_LINEFIT} package \citep{Cappellari2013}, which implements a robust linear regression algorithm that accounts for measurement uncertainties in both variables while simultaneously considering intrinsic scatter in the data. This approach effectively identifies and excludes statistical outliers. We obtained the following extinction ratios: $A_J/A_H = 1.73 \pm 0.07$, $A_H/A_{K_S} = 1.70 \pm 0.11$, and $A_J/A_{K_S} = 3.02 \pm 0.22$. The extinction correlations and their corresponding best-fit lines are displayed in Figure \ref{fig:Ax_Ay}. The linear fits yield intercepts of $-0.13 \pm 0.03$ mag, $-0.16 \pm 0.04$ mag, and $-0.48 \pm 0.05$ mag for the $A_J$ vs. $A_H$, $A_H$ vs. $A_{K_S}$, and $A_J$ vs. $A_{K_S}$ relations, respectively. These non-zero intercepts likely arise from systematic uncertainties in our extinction determination methodology, potentially reflecting systematic effects in the star count method, subtle departures from a purely linear extinction relationship at low column densities, or a combination of both factors.

	\begin{table}[h]
		\centering
		\caption{NIR extinction law parameters derived in this work and comparison with results in the literature.}
		\label{tab:ratio_result}
		\setlength{\tabcolsep}{2mm}
		\begin{tabular}{lccccc}
			\hline\hline
			Region$^a$/Literature  & $A_J/A_H$ & $A_H/A_{K_S}$ & $A_J/A_{K_S}$ & $A_J / E(J-H)$ & $A_H / E(H-K_S)$ \\
			\hline
			\multicolumn{6}{c}{This Work} \\
			\hline
			CrA (Full Cloud) & $1.73\pm0.07$ & $1.70\pm0.11$ & $3.02\pm0.22$ & $2.37\pm0.13$ & $2.43\pm0.22$ \\
			CrA-A & $1.72\pm0.13$ & $1.52\pm0.14$ & $2.75\pm0.30$ & $2.39\pm0.25$ & $2.92\pm0.52$ \\
			CrA-B & $1.78\pm0.24$ & $1.75\pm0.38$ & $3.20\pm0.72$ & $2.28\pm0.39$ & $2.33\pm0.68$ \\
			CrA-C & $1.56\pm0.19$ & $1.69\pm0.26$ & $2.68\pm0.48$ & $2.79\pm0.61$ & $2.45\pm0.55$ \\
			
			\hline
			\multicolumn{6}{c}{Literature Values} \\
			\hline
			\citet{Cardelli1989} & 1.54 & 1.56 & 2.38 & 2.85 & 2.79 \\
			\citet{Indebetouw2005} & 1.61 & 1.55 & 2.50 & 2.64 & 2.82 \\
			\citet{Nishiyama2009} & 1.75 & 1.73 & 3.02 & 2.33 & 2.37 \\
			\citet{Stead2009} & 1.77 & 1.90 & 3.37 & 2.30 & 2.11 \\
			\citet{Alonso-Garcia2017} & 1.95 & 1.95 & 3.80 & 2.05 & 2.05 \\
			\citet{Nogueras-Lara2019} & 1.84 & 1.88 & 3.46 & 2.19 & 2.14 \\
			\citet{MaizApellaniz2020} & 1.96 & 1.81 & 3.55 & 2.04 & 2.23 \\
			\citet{Hensley2023} & 1.71 & 1.60 & 2.74 & 2.41 & 2.67 \\
			\citet{Li2024b} & 1.73 & 1.71 & 2.96 & 2.37 & 2.41 \\
			\hline
		\end{tabular}
		\tablecomments{(a) ``CrA (Full Cloud)" represents measurements across the entire cloud complex, while CrA-A, CrA-B, and CrA-C designate distinct subregions within the Corona Australis molecular cloud, which are outlined in Figure \ref{fig:extinction_map}.}
	\end{table}
    
    Table \ref{tab:ratio_result} presents our measured NIR extinction ratios, with the first row showing results for the full CrA molecular cloud compared with previous studies. To quantitatively assess these literature results, we calculated unweighted means from the literature values in Table \ref{tab:ratio_result}: $A_J/A_H = 1.76 \pm 0.05$ (standard error) $\pm 0.14$ (standard deviation), $A_H/A_{K_S} = 1.74 \pm 0.05 \pm 0.15$, and $A_J/A_{K_S} = 3.09 \pm 0.17 \pm 0.49$. Our full cloud measurements ($A_J/A_H = 1.73$, $A_H/A_{K_S} = 1.70$, $A_J/A_{K_S} = 3.02$) fall well within one standard deviation of these literature means, confirming that the CrA cloud exhibits typical Galactic NIR extinction properties. The consistency between our Wolf method results and the literature values, derived from various methodological approaches, demonstrates the viability of the star count technique for determining NIR extinction ratios. The observed variations among different studies in Table \ref{tab:ratio_result} \textbf{moreso} reflect the diversity of methodological approaches, target environments, and observational constraints rather than true physical anomalies. This agreement validates that the Wolf method, despite its limitations, can reliably recover extinction ratios consistent with established Galactic values when applied to modern deep survey data.

	Additionally, the CrA molecular cloud can be divided into three distinct subregions—CrA-A, CrA-B, and CrA-C—whose spatial distributions are illustrated in Figure \ref{fig:extinction_map}. These subregions exhibit different star formation characteristics: CrA-A encompasses the Coronet cluster, CrA-B contains several starless cores, and CrA-C includes multiple dust clumps \citep{Bresnahan2018}. We perform linear fitting of the extinction ratios within each subregion, with results presented in Table \ref{tab:ratio_result}. The NIR extinction ratios in CrA-A and CrA-C are lower than the overall values for the full cloud, while CrA-B exhibits slightly higher extinction ratios. These variations across subregions may reveal fundamental differences in dust properties. However, establishing systematic relationships between these variations and specific star formation activities remains challenging. The influence of star formation on dust properties appears complex and multifaceted—beyond dust grain growth, the presence of embedded young stellar objects likely contributes to localized dust processing through outflows and radiation effects. These combined processes appear to collectively shape the observed dust properties in CrA.

    In addition to the extinction ratios, we calculate total-to-selective extinction ratios, which provide an alternative parameterization widely used in extinction studies. From our measured extinction ratios, we derive $A_J/E(J-H)$ and $A_H/E(H-K_S)$ using the relations $A_J/E(J-H) = 1/(1-1/(A_J/A_H))$ and $A_H/E(H-K_S) = 1/(1-1/(A_H/A_{K_S}))$. For the full cloud, we obtain $A_J/E(J-H) = 2.37 \pm 0.13$ and $A_H/E(H-K_S) = 2.43 \pm 0.22$. Similar calculations for the three subregions yield comparable values, with all results presented in Table \ref{tab:ratio_result}. These total-to-selective ratios derived from the Wolf method are consistent with literature values.

	\subsection{Power-law Index $\alpha$ in the NIR Extinction Law}
	
	Assuming a power-law form for the NIR extinction law, $A_\lambda\propto \lambda^{-\alpha}$,  then the extinction ratios can be converted to the power-law index $\alpha$ through:
	\begin{equation}
		\alpha_{{\lambda_1}{\lambda_2}} = \frac{\log_{10}{\frac{A_{\lambda_1}}{A_{\lambda_2}}}}{\log_{10}{\frac{\lambda_2}{\lambda_1}}},
	\end{equation}
	where $\lambda$ represents the effective wavelengths of the NIR bands. Using the VISTA effective wavelengths ($\lambda_J = 1.25\,\mu\rm m$, $\lambda_H = 1.65\,\mu\rm m$, and $\lambda_{K_S} = 2.15\,\mu\rm m$) and our measured extinction ratios, we derive $\alpha_{JH} = 2.01 \pm 0.09$, $\alpha_{JK_S} = 2.05 \pm 0.07$, and $\alpha_{HK_S} = 1.98 \pm 0.14$. The mean extinction index $\alpha \approx 2.0$ across the $JHK_S$ bands is notably higher than the standard values of 1.6-1.8 suggested by early studies \citep[e.g.,][]{Rieke1985,Draine1989,Martin1990}. Our results show excellent agreement with the value of $\sim 2.0$ reported by \citet{Wang2014} for the all-sky average and by \citet{Li2024b} for dense cloud cores. However, our derived index falls below more recent measurements that found steeper extinction laws with $\alpha$ values exceeding 2.3 \citep{Nogueras-Lara2018,Nogueras-Lara2019,Alonso-Garcia2017} and even reaching 2.64 \citep{Gosling2009}. Although previous measurements of $\alpha$ span a considerable range, our results position within the intermediate range of previously reported values.
	
	We note that the difference between $\alpha$ values in the $JH$ and $HK_S$ bands is minimal, with $\Delta\alpha = \alpha_{JH} - \alpha_{HK_S} = 0.02 \pm 0.15$, indicating no statistically significant wavelength dependence in the power-law index. This uniformity aligns with studies supporting a single power-law form across the NIR range \citep{Decleir2022,Wang2024}. Despite findings by \citet{Nogueras-Lara2019} suggesting variations in $\alpha$ between the NIR $JH$ and $HK_S$ bands, with $\alpha_{JH}$ consistently exceeding $\alpha_{HK_S}$, our results support the general assumption that a uniform power-law adequately describes the NIR extinction law across the $JHK_S$ bands.

	\subsection{Spatial Variation of the NIR Extinction law}
	
\begin{figure}[ht!]
		\hspace{0.5cm}  
		\includegraphics[width=0.7\textwidth]{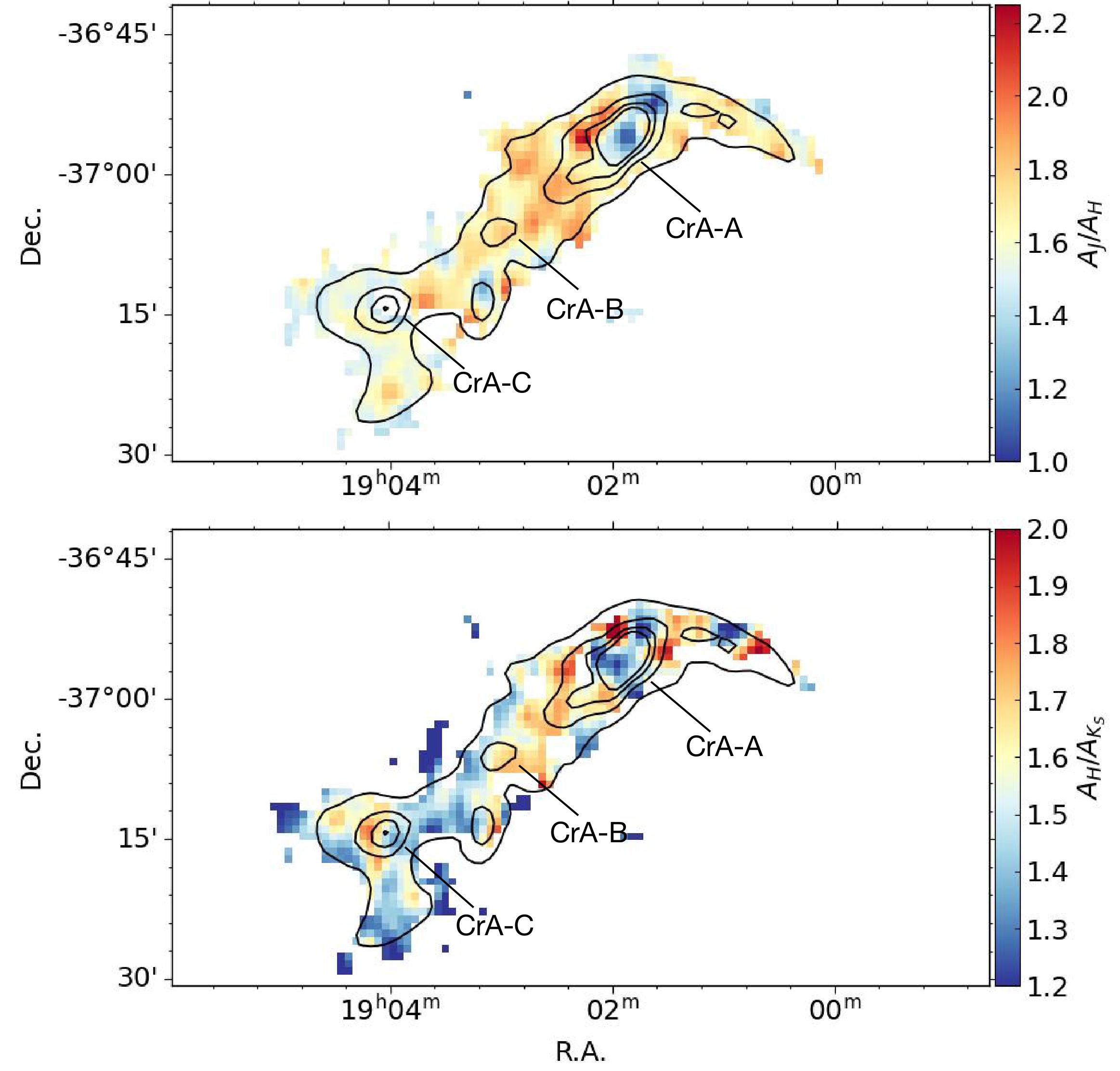}  
		\centering
		\caption{
			Spatial distribution of extinction ratios $A_J/A_H$ (top) and $A_H/A_{K_S}$ (bottom) derived from the extinction measurements presented in Figure \ref{fig:extinction_map}. Contours represent optical depth at 1\,THz ($\tau_{1\rm THz}$) for the CrA molecular cloud from \citet{Singh2022} based on Herschel observations, starting at a level of 0.001 with increments of 0.001.
		}  
		\label{fig:con}
	\end{figure}

	Figure \ref{fig:con} presents maps of extinction ratios $A_J/A_H$ and $A_H/A_{K_S}$ across the CrA molecular cloud overlaid with contours from optical depth at 1 THz from Herschel observations \citep{Singh2022}. The distributions of extinction ratios hint at the possibility of spatial variations in extinction ratios across the CrA molecular cloud, yet the broader analysis does not unequivocally confirm a robust trend. The maps show apparent variations in $A_J/A_H$ and $A_H/A_{K_S}$ ranging from approximately 1.2 to 2.0, though several factors complicate their interpretation. While some regions appear to show lower extinction ratios toward denser areas, the large uncertainties in individual measurements (e.g., $A_J/A_{K_S} = 3.20 \pm 0.72$ for CrA-B), combined with possible systematic effects from photometric cutoffs and reference field selection, prevent definitive conclusions about systematic spatial trends.
	
\begin{figure}[ht]
		\centering
		\includegraphics[width=0.5\textwidth]{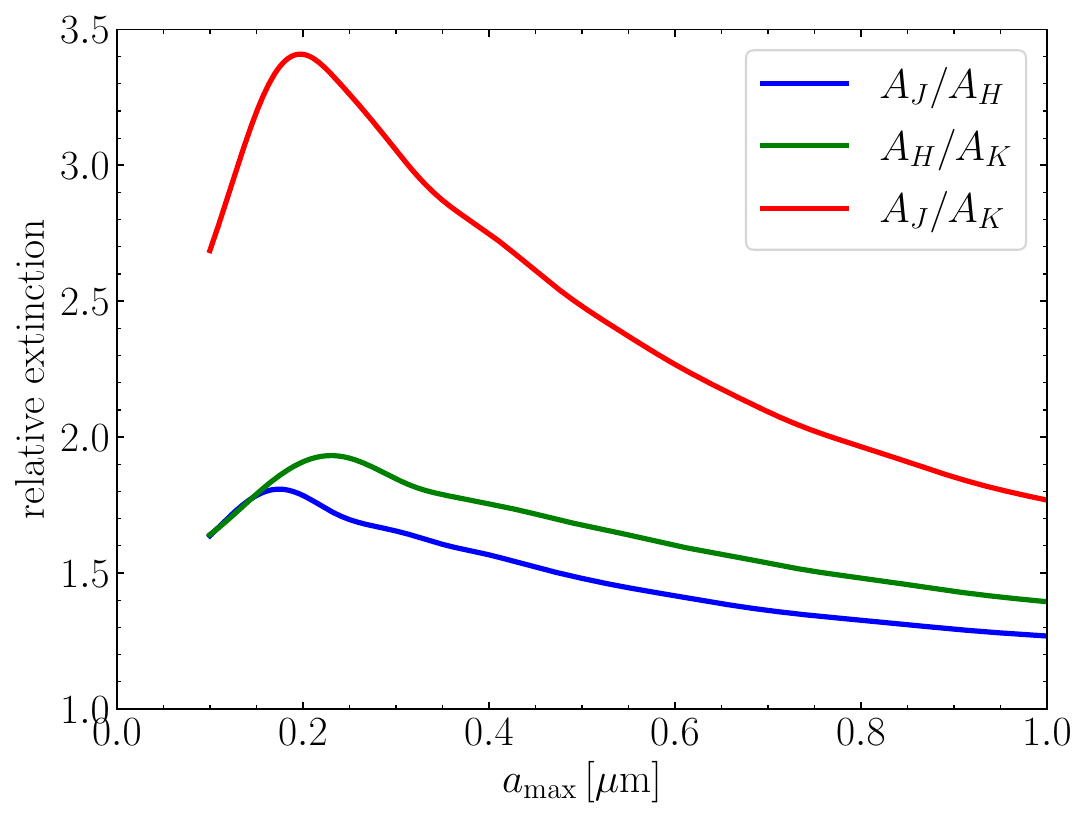} 
		\caption{Relationship between maximum grain size ($a_{\rm max}$) and NIR extinction ratios for a mixture of silicate and graphite grains with a power-law size distribution (${\rm d}n/{\rm d}a\propto a^{-3.5}$) and minimum grain size $a_{\rm min} = 0.005\,\mu\rm m$. Three extinction ratios are shown: $A_J/A_H$ (blue), $A_H/A_{K_S}$ (green), and $A_J/A_{K_S}$ (red). All ratios systematically decrease as maximum grain size increases from 0.1 to 1.0\,$\mu$m, providing a theoretical framework for interpreting observed variations in extinction properties across the CrA cloud. }
		\label{fig:relative_extinction_model}
	\end{figure}

	Despite these limitations, suppose such variations are indeed real. The apparent tendency for lower extinction ratios in dense regions could potentially be explained by grain growth processes. We explore this possibility using theoretical models. First, we consider a mixture of silicate and graphite grains \citep{Draine1989} following a power-law size distribution $dn/da\propto a^{-3.5}$, with a fixed minimum grain size $a_{\rm min}=0.005\,\mu$m and a variable maximum radius $a_{\rm max}$. Figure \ref{fig:relative_extinction_model} shows that these models predict decreasing extinction ratios with increasing maximum grain size when $a_{\rm max}\gtrsim0.2\,\mu\rm m$. If the hints of variation in Figure \ref{fig:con} represent real physical changes, they could potentially indicate grain growth in dense environments, though this remains speculative.
	
\begin{figure}[ht]
		\centering
		\includegraphics[width=0.7\textwidth]{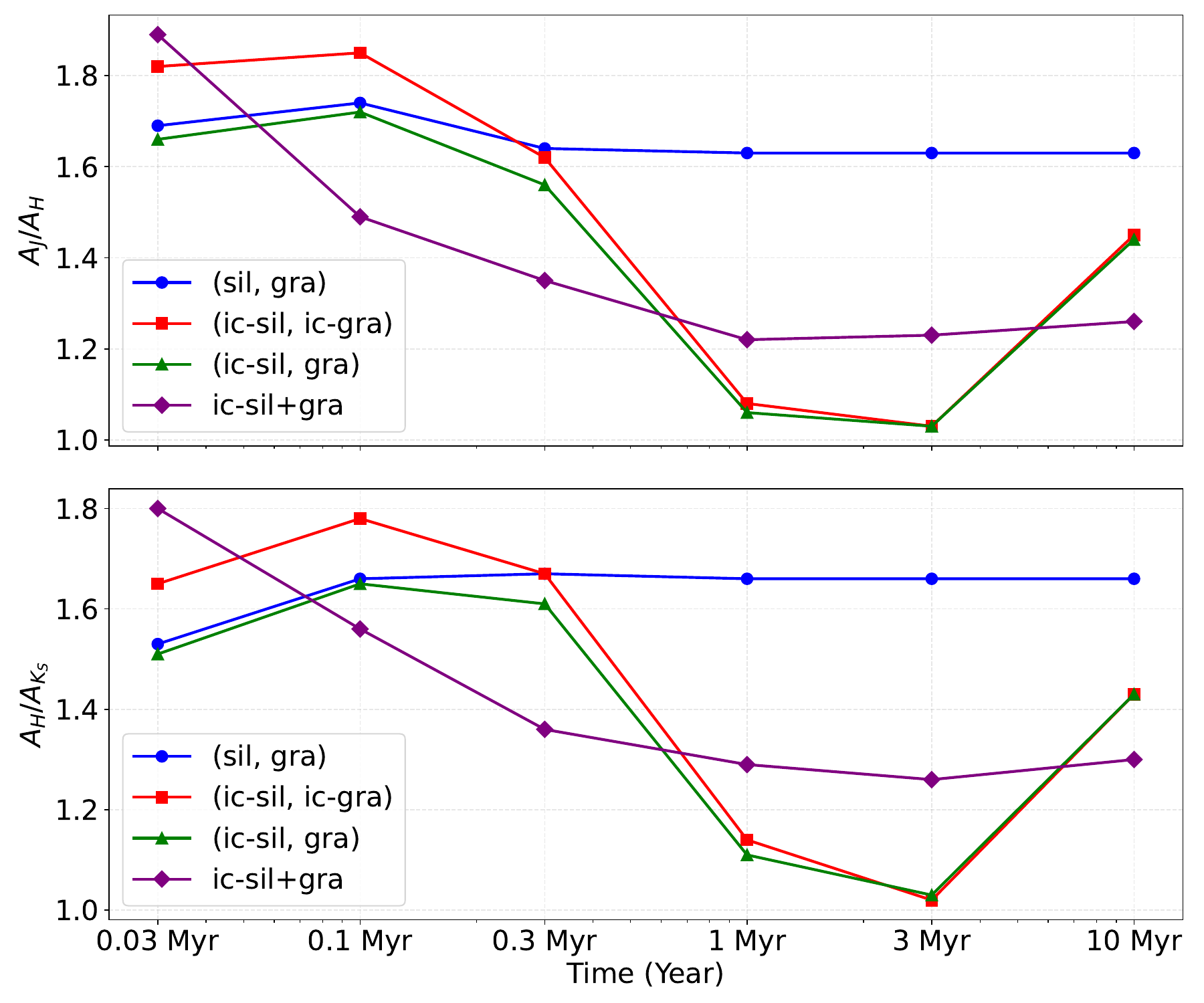} 
		\caption{Evolution of NIR extinction ratios $A_J/A_H$ (top) and $A_H/A_{K_S}$ (bottom) over time for four dust models based on \citep{Ormel2011}: silicate and graphite without ice (blue), ice-coated silicate and graphite (red), ice-coated silicate with uncoated graphite (green), and ice-coated silicate with graphite inclusions (purple). While ice-free compositions maintain stable ratios, ice-coated models show significant decreases in extinction ratios over timescales of 0.1\text{--}10\,Myr, demonstrating how ice mantles affect grain growth and observable extinction properties. }
		\label{fig:ormel_model_ext}
	\end{figure}

	We also examine the grain growth models of \citet{Ormel2011}, which incorporate various dust compositions and ice mantle effects. Figure \ref{fig:ormel_model_ext} illustrates how different dust models might evolve over time. The models suggest that ice-coated grains could potentially produce lower extinction ratios through enhanced growth efficiency. If our tentatively observed variations are real, the range of values ($A_J/A_H\sim$1.2--2.0) could potentially be explained by different evolutionary stages of dust processing, with ice mantle formation playing a possible role. This also aligns with spectroscopic studies revealing abundant ice features in dense cores \citep{Boogert2015} and with millimeter observations showing enhanced emissivity associated with evolved grain populations \citep{Ysard2016}.
	However, we stress that these model comparisons are exploratory and cannot confirm whether the apparent spatial variations represent genuine dust evolution or methodological artifacts.

	If future studies with improved methodologies and observations (i.e. James Webb Space Telescope) confirm spatial variations in NIR extinction properties, this would have implications for extinction corrections in star-forming regions. However, at present, we can only note that our data suggest the possibility of environmentally dependent extinction properties that merit further investigation. The potential connection between extinction variations and dust evolution remains an open question requiring more robust observational constraints.

	\section{Summary}\label{sec:Sum}
	
	In this study, we assess the viability of the Wolf method for constraining NIR extinction laws in the CrA molecular cloud using deep $JHK_S$ observations from the VISIONS survey. Our principal findings include:
	
	\begin{enumerate}
		\item{We construct extinction maps at 1 arcmin resolution in $J$, $H$, and $K_S$ bands, revealing the density structure of the cloud with extinction values reaching $A_J$ = 6.8 mag in the densest regions. The methodology demonstrates both the potential and limitations of applying the Wolf approach to modern survey data.}
		
		\item{We derive NIR extinction ratios of $A_J/A_H = 1.73 \pm 0.07$, $A_H/A_{K_S} = 1.70 \pm 0.11$, and $A_J/A_{K_S} = 3.02 \pm 0.22$. These values are quantitatively consistent with Galactic literature means, falling well within one standard deviation of typical interstellar extinction properties. The corresponding power-law indices ($\alpha_{JH} = 2.01 \pm 0.09$, $\alpha_{HK_S} = 1.98 \pm 0.14$, and $\alpha_{JK_S} = 2.05 \pm 0.07$) confirm standard NIR power-law behavior.}
		
		\item{Our extinction ratio maps hint at possible spatial variations across the cloud, though the broader analysis does not unequivocally confirm systematic trends. The apparent variations, if real, could potentially indicate dust evolution in dense environments. However, concerns persist regarding the impact of photometric cutoffs and reference field selection on these tentative results. Further research with refined methodologies is required to verify whether such variations represent genuine physical trends or methodological artifacts.}
	\end{enumerate}

	\section*{Acknowledgments}
	We thank the anonymous referee for his/her insightful comments and suggestions. This work is supported by the National Natural Science Foundation of China (NSFC) through project Nos. 12403026 and 12473024, the National Key R\&D program of China 2022YFA1603102. X.C. thanks to Guangdong Province Universities and Colleges Pearl River Scholar Funded Scheme (2019).

	\bibliography{ref}{}
	\bibliographystyle{aasjournal}
	
	\clearpage
\end{CJK}
\end{document}